\documentclass[prb,twocolumn,aps,showpacs]{revtex4}
\usepackage{graphicx,epsfig}% complex graphics
\usepackage{bm}% bold math\textbf{}
\usepackage{amssymb} %math symbols
\begin{document}

\title{Phase diagram of weak-magnetic-field quantum Hall transition
quantified from classical percolation}

\author{M. Ortu\~{n}o$^{1}$, A. M. Somoza$^{1}$, V. V. Mkhitaryan$^{2}$, and M. E. Raikh$^{2}$}

\affiliation{$^{1}$Departamento de F\'{i}sica - CIOyN,
Universidad de Murcia, Murcia 30.071, Spain\\
$^{2}$Department of Physics, University of Utah, Salt Lake City,
UT 84112, USA}

\begin{abstract}
We consider magnetotransport in high-mobility 2D electron gas,
$\sigma_{xx}\gg1$, in a non-quantizing magnetic field. We employ a
weakly chiral network model to test numerically the prediction of
the scaling theory that the transition from an Anderson to a
quantum Hall insulator takes place when the Drude value of the
non-diagonal conductivity, $\sigma_{xy}$, is equal to $1/2$ (in
the units of $e^2/h$). The weaker is the magnetic field the harder
it is to locate a delocalization transition using quantum
simulations. The main idea of the present study is that the {\it
position} of the transition does not change when a strong {\it
local} inhomogeneity is introduced. Since the strong inhomogeneity
suppresses interference, transport reduces to classical
percolation. We show that the corresponding percolation problem is
bond percolation over two sublattices coupled to each other by
random bonds. Simulation of this percolation allows to access the
domain of very weak magnetic fields. Simulation results confirm
the criterion $\sigma_{xy}=1/2$ for values $\sigma_{xx}\sim 10$,
where they agree with earlier quantum simulation results. However
for larger $\sigma_{xx}$ we find that the transition boundary is
described by $\sigma_{xy}\sim\sigma_{xx}^\kappa$ with
$\kappa\approx 0.5$, i.e., the transition takes place at higher
magnetic fields. The strong inhomogeneity limit of
magnetotransport in the presence of a random magnetic field,
pertinent to composite fermions, corresponds to a different
percolation problem. In this limit we find for the delocalization
transition boundary $\sigma_{xy}\sim\sigma_{xx}^{0.6}$.
\end{abstract}
\pacs{72.15.Rn; 73.20.Fz; 73.43.-f}

\maketitle

\section{Introduction}

Anderson localization is a {\it single-particle} phenomenon.
Nevertheless, the scaling theory of localization \cite{AALR} which
yields a profound prediction, full localization of all states in
two dimensions, was formulated in terms of conductivity of {\it
electron gas}, $\sigma$. Similarly, the extension \cite{pru84} of
the 2D scaling theory to a finite magnetic field is formulated in
terms of components, $\sigma_{xx}$ and $\sigma_{xy}$, of the
conductivity tensor of electron gas. Scaling equations describing
the evolution of these components with the sample size, $L$, have
the form
\begin{eqnarray}
\label{pru1} &&\frac{\partial\sigma_{xx}}{\partial\ln
L}=-\frac{1}{2\pi^2\sigma_{xx}}-
\sigma_{xx}^2{\cal D}e^{-2\pi\sigma_{xx}}\cos(2\pi\sigma_{xy}),\\
&&\frac{\partial\sigma_{xy}}{\partial\ln L}= -\sigma_{xx}^2{\cal
D}e^{-2\pi\sigma_{xx}}\sin(2\pi\sigma_{xy}),\label{pru2}
\end{eqnarray}
where ${\cal D}$ is a dimensionless constant. Drude values of
$\sigma_{xx}$ and $\sigma_{xy}$ at size, $L$, of the order of mean
free path, $l$, are given by
\begin{equation}
\label{Drude} \sigma_{xx}{\big |}_{L\sim
l}=\frac{\sigma_0}{1+(\omega_c\tau)^2},\qquad \sigma_{xy}{\big
|}_{L\sim l}=\frac{\sigma_0\,(\omega_c\tau)}{1+(\omega_c\tau)^2},
%where $\sigma_0 \propto E_n$ is the dimensionless conductance at
%$\omega_c=0$,
\end{equation}
where $\sigma_0=k_{\scriptscriptstyle F}l$, $k_{\scriptscriptstyle
F}$ is the Fermi wavevector, $\omega_c$ is the cyclotron
frequency, and $\tau$ is the scattering time. These values serve
as initial conditions to Eqs. (\ref{pru1}) and (\ref{pru2}). Fixed
points, $\sigma_{xy}=n+1/2$, at which $\sigma_{xx}$ is finite,
determine the energies of delocalized states,
\begin{equation}
\label{positions} E_n =\hbar\omega_c\left(n+\frac{1}{2}\right)
\left[1+\frac1{(\omega_c\tau)^{2}}\right].
\end{equation}
%%%%%%%%%%%%%%%%%%%
\begin{figure}[t]
\centerline{\includegraphics[width=90mm,angle=0,clip]{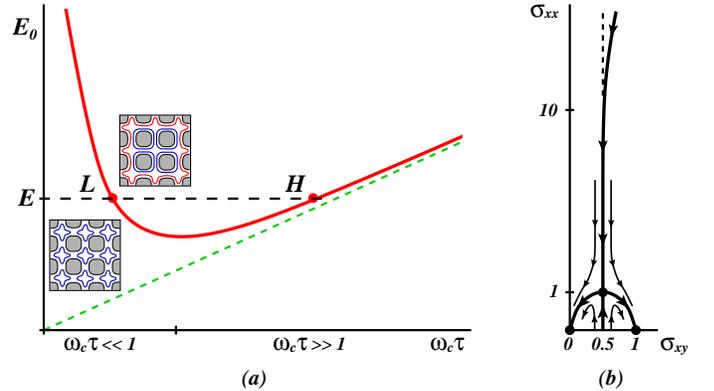}}
\caption{(Color online) (a) Energy position of delocalized state,
$E_0$, versus magnetic field, $\omega_c$, as predicted by Eq.
(\ref{positions}).
%At a given $E$, delocalization transition occurs at low field
%(point $L$) and at high field (point $H$).
The curve, $E_0(\omega_c)$, separates the phases with quantized
Hall conductivities, $\sigma_{xy}=0$ and $\sigma_{xy}=1$. Cartoons
illustrate electron trajectories with restricted geometry in both
phases; the edge state (red) is present in the upper cartoon and
absent in the lower cartoon. (b) The predicted modification of the
form of the flow diagram \cite{Khmelnitskii} is illustrated
schematically.} \label{levit}
\end{figure}
%%%%%%%%%%%%%%%%%%

The most nontrivial consequence of Eq. (\ref{positions}) is that
it predicts levitation of delocalized states in weak magnetic
fields $\omega_c\tau\ll1$, see Fig. \ref{levit}a. In such fields
it takes the form $E_n =(n+\frac12)\hbar/\omega_c\tau^2$. In
physical terms this means that a high-mobility electron gas with
zero-field Drude conductivity $\sigma_{xx}=E_{\scriptscriptstyle
F}\tau/\hbar\gg1$ exhibits a very strong sensitivity to a weak
magnetic field,
\begin{equation}
\label{increase}
\omega_c\tau\sim \hbar/E_{\scriptscriptstyle
F}\tau,
\end{equation}
as the temperature is decreased and quantum interference effects
become important. Phenomenon of levitation was predicted by
Khmelnitskii \cite{Khmelnitskii} even before Eqs. (\ref{pru1}) and
(\ref{pru2}) were put forward (see also Ref. \onlinecite{Laugh}).
Subsequently, it was also observed experimentally by several
groups \cite{Jiang92, Jiang93, Jiang93PRL, Jiang95PRL, Jiang95,
Wang94, Hughes94, ShaharLev}.
%In weak field, the condition $\sigma_{xy}=1$ translates into
%\begin{equation}
%\label{wfl} k_{\scriptscriptstyle F}l
%\omega_c\tau=l^2/\lambda^2=Bl^2/\phi_0=1,
%\end{equation}
%i.e., a square with the side $l$ is pierced by a flux quantum. In
%strong field, the same condition yields $k_{\scriptscriptstyle F}l
%/\omega_c\tau=E_{\scriptscriptstyle F}/\hbar \omega_c=1$, which
%means that there is one Landau level under the Fermi energy.

\subsection{Physical interpretation of Eq. (\ref{pru1})}

The starting point in derivation of scaling equations (\ref{pru1})
and (\ref{pru2}) was a $\sigma$-model with topological term
\cite{pru84}. It is desirable to understand physical processes
underlying these equations. The first term Eq. (\ref{pru1}) comes
from Aharonov-Bohm {\it phase} action of the magnetic field. It
describes that two paths corresponding to the same scatterers but
different sequences of scattering events interfere even in the
presence of the Aharonov-Bohm phases. The interpretation of the
second term in Eq. (\ref{pru1}) is transparent in the limit of
classically strong magnetic field, $\omega_c\tau>1$, where
$\cos(2\pi\sigma_{xy})$ assumes the form $\cos(2\pi
E_{\scriptscriptstyle F}/\hbar\omega_c)$, which is simply the
field-induced modulation of the density of states. The origin of
modulation is the emergence of Landau levels.
%Discreteness of Landau levels even in the presence of disorder is
%implied in the second term Eq. (\ref{pru1}).
On the other hand, Landau levels reflect the {\it orbital} action
of the magnetic field, i.e., the fact that, with a certain
probability, an electron can complete a Larmour circle with
radius, $R_{\scriptscriptstyle L}$, without being scattered away
by disorder. Thus the interpretation of the right-hand side of Eq.
(\ref{pru1}) is that the phase and orbital actions of the magnetic
field compete with each other.

Unlike strong fields, the interpretation of the cosine term in Eq.
(\ref{pru1}) in weak fields, $\omega_c\tau\ll1$, is
%quite mysterious.
much less transparent. In this limit we have
$\sigma_{xy}=\sigma_0\omega_c\tau$ in the argument of cosine. The
cosine term can be also rewritten as
%the oscillating term assumes the form
\begin{equation}\label{wf}
\cos[2\pi (k_{\scriptscriptstyle F}l)
(\omega_c\tau)]=\cos\left(\frac{2\pi l^2}{l_B^2}\right)
=\cos\left(\frac{2\pi Bl^2}{\Phi_0}\right),
\end{equation}
where $l_B$ is the magnetic length, and $\Phi_0$ is the flux
quantum. For comparison, in the strong-field limit, the cosine
term can be cast in the form
\begin{equation}\label{wfcos}
\cos(2\pi BR_{\scriptscriptstyle L}^2/\Phi_0).
\end{equation}
Comparing this expression to the last cosine in Eq. (\ref{wf})
suggests that in weak fields the role of the Larmour radius is
taken by the mean free path $l$. Note that $l$ does not depend on
magnetic field. Then the following question arises: what physics
causes the orbital action of the magnetic field to manifest itself
in the scaling equations in the weak-field limit? A possible way
to unveil the orbital action is to adopt a cartoon picture where
an electron moves not in a random potential but rather in a {\it
periodic} background, say, on a quadratic lattice, as in seminal
paper Ref. \onlinecite{TKNN}. Then we have to assume that the
lattice constant, $l$, is {\it set by disorder}. In this cartoon
the orbital action will be encoded into the structure of the Bloch
wavefunctions of electrons. It is the structure of the Bloch
wavefunctions that leads to edge states in the presence of
boundaries \cite{TKNN}. Note that the structure of the Bloch
wavefunctions in a magnetic field depends crucially on the number
of flux quanta through the unit cell, which, upon identifying $l$
with the lattice constant, is the argument of the cosine in Eq.
(\ref{wf}). Then the factor, $\exp[-2\pi k_{\scriptscriptstyle
F}l]$, in front of the cosine in Eq. (\ref{pru1}) has a meaning of
degree to which a realistic random potential can be viewed as a
periodic. Indeed, this factor can be interpreted as a probability
for a realistic diffusive electron to execute the {\it same} loop
of length $\sim l$ more than once.

Obviously, realistic disordered system does not have any built-in
spatial periodic structure. In view of the lack of a transparent
interpretation of topological term in the weak-magnetic-field
limit, it is important to check numerically whether or not some
discrete value of the magnetic field of the order of $B\sim
\Phi_0/l^2$ causes delocalization transition and formation of edge
state in a random potential. This was a subject of the papers Ref.
\onlinecite{our}. In these papers, a network model describing a
weakly chiral electron motion was introduced. The position of the
quantum delocalization transition was established from the
conventional transfer-matrix simulation of the transmission of the
network. It was demonstrated that up to $k_{\scriptscriptstyle
F}l\sim10$ the above estimate for the transition field applies.
However quantum simulations become progressively complex in the
limit of vanishing field.

\subsection{Delocalization transition with strong spatial inhomogeneity}

In the present paper we establish the position of delocalization
transition indirectly. The underlying idea of our approach is that
when a strong spatial inhomogeneity is introduced into the quantum
network, interference effects become progressively irrelevant, in
the sense, that {\it amplitudes} of each two interfering paths
typically differ strongly. Then the problem of the transport
through a network reduces to the {\it classical percolation}. Most
importantly, while the inhomogeneity-induced suppression of
quantum interference leads to a strong reduction of the
localization radius, the {\it position} of the transition remains
unchanged. At the same time, classical simulations can be extended
to much weaker magnetic field. The main outcome of our simulations
is that for very weak fields (very small $\omega_c\tau$) or high
electron energies (very large $k_{\scriptscriptstyle F}l$) the
transition field is higher than $\Phi_0/l^2$, namely
\begin{equation}
\label{newscale} B\sim \frac{\Phi_0}{l^2} (k_{\scriptscriptstyle
F}l)^{\kappa},
\end{equation}
where $\kappa$ is close to $1/2$. In terms of the flow diagram of
the quantum Hall effect \cite{Khmelnitskii} the result Eq.
(\ref{newscale}) translates into the prediction that for
$\sigma_{xx}>10$ the vertical flow line in Fig.~\ref{levit}b
deviates from $\sigma_{xy}=1/2$ to the right.

In the present paper we also study levitation of delocalized
states in a vanishing {\it average} magnetic field, but in the
presence of a strongly fluctuating {\it random} magnetic field.
There is a notion that electron density variations near the
half-filling, $\nu=1/2$, of the lowest Landau level reduces to
random magnetic field acting on composite fermions
\cite{Jain,HLR}. It is also possible to realize an inhomogeneous
magnetic field, acting on 2D electrons, artificially
\cite{Geim90,bending90,Geim92,
Geim94,smith94,mancoff95,gusev96,gusev00,rushforth04}. Different
aspects of electron motion in random magnetic fields have been
studied theoretically in
Refs.~\onlinecite{Chalker94,Chalker94',Aronov94,chklovskii94,chklovskii95,
Falko94,Khveshchenko96,Simons99,Shelankov00,Mirlin1,Mirlin2,Mirlin3,baranger01,efetov04}.
%In parallel to magnetotransport of electrons in weak magnetic
%field we also study magnetotransport of composite fermions at
%filling factors very close to $\nu=1/2$.

Fractional quantum Hall transitions can be associated with
quantization of cyclotron orbits of composite fermions. In this
sense, fractional quantum Hall transitions are the counterparts of
delocalization transitions of electrons. Then the question arises:
whether a delocalization transition for electrons at vanishing
magnetic fields has its counterpart for composite fermions at
vanishing $|\nu-1/2|$. At such filling factors, composite fermion
"feels" very weak average magnetic field. On the other hand, local
fluctuations of electron density give rise to a very strong {\it
random} magnetic field, acting on composite fermion. For this
situation we reduce the description of magnetotransport to a
different percolation problem. For critical values of filling
factors we obtain $|\nu-1/2|\sim(k_{\scriptscriptstyle
F}l)^{-0.4}$.

\section{Weakly-chiral network model}

\subsection{Description}

\begin{figure}[t]
\centerline{\includegraphics[width=80mm,angle=0,clip]{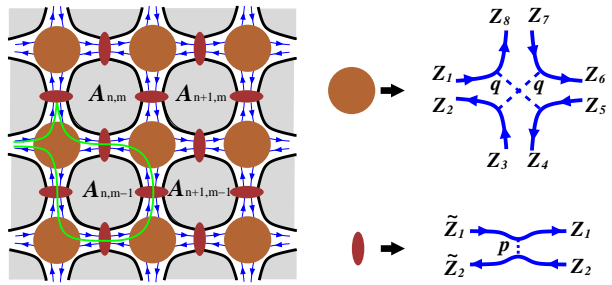}}
\caption{(Color online) Left: Restricted electron motion over
point contacts and bend-junctions is illustrated; $A_{n,m}$ are
the centers of forbidden regions. Green line shows a minimal loop
that can be traversed in clockwise and anticlockwise directions.
Right: Scattering matrices at the node and at the link. }
\label{pqmodel}
\end{figure}

For completeness we remind the construction of the network model
introduced in Refs. \onlinecite{our} to describe quantum electron
motion in a weak magnetic field. This construction is illustrated
in Fig.~\ref{pqmodel} and consists of three steps:

({\it i}) We restrict electron motion by introducing forbidden
regions, $A_{nm}$, (grey areas in Fig.~\ref{pqmodel}), which are
not accessible for electron. Then electron moves in both
directions along the links, which are the white regions,
separating $A_{nm}$. The links join each other at the nodes, shown
in Fig.~\ref{pqmodel} with brown full circles.

({\it ii}) We forbid forward and backward scattering at the nodes.
This allows to parameterize the node scattering matrix, $S_q$, by
a single parameter, $q$, as follows:
\begin{equation}
\label{Matrixq} \left(
\begin{array}{c}
Z_2\\
Z_4\\
Z_6\\
Z_8
\end{array}\right)=\left(
\begin{array}{cccc}
0&{\scriptstyle-\sqrt{1-q}}&0&{\scriptstyle -\sqrt{q}}\\
{\scriptstyle \sqrt{q}}&0&{\scriptstyle \sqrt{1-q}}&0\\
0&{\scriptstyle -\sqrt{q}}&0&{\scriptstyle \sqrt{1-q}}\\
{\scriptstyle \sqrt{1-q}}&0&{\scriptstyle -\sqrt{q}}&0
\end{array}\right)\left(
\begin{array}{c}
Z_1\\
Z_3\\
Z_5\\
Z_7
\end{array}\right),
\end{equation}
where $Z_i$ are the amplitudes of incoming and outgoing waves, see
Fig.~\ref{pqmodel}.

({\it iii}) We incorporate  backscattering of electron moving
along the link. The probability of backscattering is $p$, so that
the corresponding scattering matrix, $S_p$, has the form
\begin{equation}
\label{Matrixp} \left(
\begin{array}{c}
Z_1\\
\tilde{Z}_2
\end{array}\right)=\left(\begin{array}{cc}
{\scriptstyle \sqrt{1-p}}&{\scriptstyle \sqrt{p}}\\
{\scriptstyle -\sqrt{p}}&{\scriptstyle \sqrt{1-p}}
\end{array}\right) \left(
\begin{array}{c}
\tilde{Z}_1\\
Z_2
\end{array}\right),
\end{equation}
where $\tilde{Z}_1$ and $Z_2$ are amplitudes of incident waves,
whereas $Z_1$ and $\tilde{Z}_2$ are amplitudes of reflected waves,
see Fig.~\ref{pqmodel}.
%%%%%%%%%%%%%%%%%%%
%%%%%%%%%%%%%%%%%%

To establish a correspondence with physical parameters, we
identify the lattice constant with the mean free path, $l$. To
relate the parameter $q$ with the magnetic field, one can express,
following Refs. \onlinecite{Ravenhall, Baranger, Beenakker}, the
Hall resistance of the node, $R_H$, via the elements of matrix
$S_q$, Eq. (\ref{Matrixq}):
\begin{equation}\label{HRq}
R_H=\frac{2q-1}{q^2+(1-q)^2}.
\end{equation}
In the absence of magnetic field $R_H$ vanishes, indicating that
$(1/2-q)$ is a measure of magnetic field, which is also the degree
of preferential scattering to the left over scattering to the
right. For a realistic electron moving a distance $l$ in a
magnetic field, this degree is $\omega_c\tau$, thus allowing the
following identification:
\begin{equation}\label{qiden}
\frac12-q=\omega_c\tau.
\end{equation}
Note that electron motion over links and scattering at nodes in
Fig.~\ref{pqmodel} models adequately the transport in the
Boltzmann limit even at $p=0$, i.e., without backscattering.
However, a specifics of the diffusive motion with $p=0$ is that it
{\it does not} allow quantum weak localization corrections.
Indeed, weak localization corrections originate from the
trajectories on the network for which an electron, starting from a
certain link, returns to the same link with {\it opposite}
direction of velocity (coherent backscattering). At $p=0$ the
electron still can return to the same link, e.g., by encircling
one forbidden region, but its velocity will be {\it the same} as
the initial velocity. Finite $p$ gives rise to weak localization.
An example of an elementary loop providing coherent backscattering
is shown in Fig.~\ref{pqmodel}. The probability of this loop is
\begin{equation}\label{backprob}
{\cal P}=p(1-p)^4\bigl[q(1-q)\bigr]^3.
\end{equation}
For realistic electron the return probability is
$(k_{\scriptscriptstyle F}l)^{-1}$. This allows us to identify the
parameter $p$ as
\begin{equation}\label{piden}
p=\frac1{k_{\scriptscriptstyle F}l}=\frac1{\sigma_0}.
\end{equation}
To conclude the construction of the quantum network, we assume as
usually that random phases are accumulated in course of
propagation along the links. This convention is non-trivial in the
weak field limit. Indeed, as we discussed in the Introduction, the
delocalization transition is expected when the magnetic flux
through a plaquette is of the order of flux quantum, $\Phi_0$. We
will return to this point in Section IV D.
%%%%%%%%%%%%%%%%%%%
\begin{figure}[t]
\centerline{\includegraphics[width=50mm,angle=0,clip]{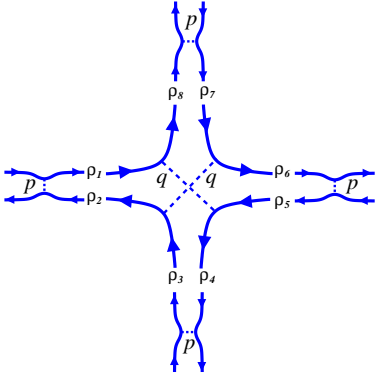}}
\caption{(Color online) Boltzmann transport on the p-q network.
Rate equations Eq. (\ref{ratesys}) relate the probabilities,
$\rho_i(m,n;t)$, to find electron on the corresponding half-link
adjacent to the node with coordinates $(m,n)$ and at time
instances $t$ and $t+\tau$.} \label{boltz}
\end{figure}
%%%%%%%%%%%%%%%%%%

The presence of two types of scattering processes, on the links
and at the nodes, makes the Boltzmann description of transport
more complex. To develop this description, we turn to Fig.
\ref{boltz}. It illustrates that the adequate variables to
describe the Boltzmann transport are the probabilities, $\rho_i$,
$i=1,..,8$, to find an electron on corresponding "half-link". In
these variables, the closed set of rate equations reads
\begin{eqnarray}\label{ratesys}
&&\rho_1(m,n;t+\tau)=[1-p]\rho_6(m-1,n;t)+p\rho_2(m,n;t),\nonumber\\
&&\rho_2(m,n;t+\tau)=[1-q]\rho_3(m,n;t)+q\rho_7(m,n;t),\nonumber\\
&&\rho_3(m,n;t+\tau)=[1-p]\rho_8(m,n-1;t)+p\rho_4(m,n;t),\nonumber\\
&&\rho_4(m,n;t+\tau)=[1-q]\rho_5(m,n;t)+q\rho_1(m,n;t),\nonumber\\
&&\rho_5(m,n;t+\tau)=[1-p]\rho_2(m+1,n;t)+p\rho_6(m,n;t),\nonumber\\
&&\rho_6(m,n;t+\tau)=[1-q]\rho_7(m,n;t)+q\rho_3(m,n;t),\nonumber\\
&&\rho_7(m,n;t+\tau)=[1-p]\rho_4(m,n+1;t)+p\rho_8(m,n;t),\nonumber\\
&&\rho_8(m,n;t+\tau)=[1-q]\rho_1(m,n;t)+q\rho_5(m,n;t).\nonumber\\
&&\,
\end{eqnarray}
Performing Fourier transform in time and coordinate domains and
taking the limit of small momenta, $k$, and frequencies,
$\omega_k$, we find a diffusive mode $-i\omega_k=Dk^2$, where $D$
is given by
\begin{equation}\label{diffcoeff}
D=\left(\frac{l^2}{4\tau}\right)\frac{1-p}8\frac{1+(2q-1)^2(2p-1)}
{1+(2q-1)^2(2p-1)^2}.
\end{equation}
As discussed above, the diffusion coefficient is finite even at
$p=0$, except in the "strong-field" limits, $q=1$ and $q=0$, where
the electron circulates around forbidden regions, clockwise and
anti-clockwise, respectively. In these limits, for small $p$ the
diffusion coefficient is proportional to $p$. From Eq.
(\ref{diffcoeff}) we also see that $D\rightarrow0$ in the
strong-scattering limit, $p\rightarrow1$, as could be expected. In
the limit of weak magnetic field, $(1/2-q)\ll1$, and high
mobility, $p\ll1$, we have
\begin{equation}\label{diffcoeff}
D=\left(\frac{l^2}{32\tau}\right)\left[1-p-8\left(\frac12-q\right)^2\right].
\end{equation}
The fact that the magnetic-field correction to $D$ is
$\sim(\omega_c\tau)^2$ is generic for classical magnetotransport.
On the other hand, the negative classical correction to $D$ due to
finite $p$ is model-specific, since $p$ was incorporated to
capture interference effects. The meaning of the prefactor,
$l^2/\tau$, which emerges from the system Eq. (\ref{ratesys}) in
the course of Fourier transform, is that the electron travels the
distance of the mean free path, $l$, during scattering time,
$\tau$. Correct, within a number, prefactor and magnetic field
dependencies of $D$ indicates that the network model captures
properly the magnetotransport in high-mobility electron gas in the
Boltzmann limit.
%%%%%%%%%%%%%%%%%%%
\begin{figure}[t]
\centerline{\includegraphics[width=55mm,angle=0,clip]{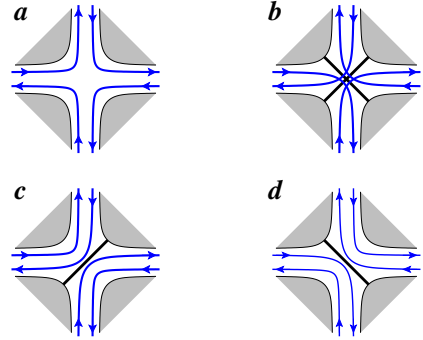}}
\caption{(Color online) Definition of the q-bonds. Scattering
scenarios (a), (b), (c), and (d) correspond to the absence of both
q-bonds, presence of both q-bonds, presence of one right-diagonal
q-bond, and presence of one left-diagonal q-bond, respectively.}
\label{pqbonds}
\end{figure}
%%%%%%%%%%%%%%%%%%

According to the scaling theory of localization, the knowledge of
the Boltzmann transport coefficient should be sufficient to
predict the position, Eq. (\ref{positions}), of the {\it quantum}
delocalization transitions, which in the limit of weak fields
takes the form $E_0 \sim\hbar/\omega_c\tau^2$. In the language of
the network model this translates into the linear dependence,
\begin{equation}
\label{predbound} p\sim\frac12-q.
\end{equation}
Whether or not this prediction is valid can be established only by
quantum numerical simulations. Especially important is the limit,
$q\rightarrow1/2$, which corresponds to vanishing magnetic fields
where a strong levitation is expected. Unfortunately this limit is
the hardest to simulate. This is because the localization radius
to the left and to the right of the delocalization transition is
huge, $\ln(\xi/l)=\pi^2\sigma_0^2\sim\pi^2/p^2$. This was a
limitation of the quantum simulations reported in Refs.
\onlinecite{our}, where the smallest value of $p$ was $p=0.1$.

\subsection{From quantum delocalization to classical percolation}

There is another, indirect, way to find the critical $p-q$
boundary, bypassing quantum simulations, namely, to take the limit
of strong disorder. By a limit of strong disorder we mean that
{\it local} values $p_i$ and $q_i$ are strongly spread around
averages $p$ and $q$ with distributions
\begin{eqnarray}\label{disp}
&&f(p_i)=p\,\delta(1-p_i)+(1-p)\delta(p_i),\\
&&f(q_j)=q\,\delta(1-q_j)+(1-q)\delta(q_j). \label{disq}
\end{eqnarray}
Unlike the quantum case, where $p_i$ and $q_i$ were the same for
all links and nodes, with distribution Eq. (\ref{disp}) scatterers
on the links reflect {\it fully} in  $p$ percent of the cases, and
transmits fully in the rest $(1-p)$ percent of cases. Similarly,
according to Eq. (\ref{disq}), the nodes deflect only to the right
in $q^2$ percent of the cases, deflect only to the left in
$(1-q)^2$ percent of the cases; in the remaining $2q(1-q)$ percent
of the cases the deflection takes place both to the left and to
the right depending on the incoming channel, see
Fig.~\ref{pqbonds}. The advantage of the strong disorder limit is
that the quantum interference effects are irrelevant. The simplest
way to see this is to turn to the elementary interference process
illustrated in Fig. \ref{pqmodel}. If return to the origin is
allowed for the clockwise direction, then it is forbidden for the
anti-clockwise direction since $q_i(1-q_i)$ is zero in the
strong-disorder limit.

In the absence of interference the transport reduces to the
classical bond percolation problem. The reduction is achieved by
replacing scattering matrices Eqs. (\ref{Matrixq}) and
(\ref{Matrixp}) by bonds according to the following rules:

({\it i}) The realization in which $p_i=1$ corresponds to
quantum-mechanical reflection of incoming waves from all
directions. In the language of percolation this configuration
corresponds to a {\it bond} installed between the neighboring
forbidden regions, $A_{n,m}$ and $A_{n+1,m}$, i.e., horizontal
bond in Fig. \ref{pqmodel}. Below we will refer to this bond as a
p-bond. For configurations with $p_i=0$ the p-bond between the
neighboring forbidden regions $A_{n,m}$ and $A_{n+1,m}$ is absent.

({\it ii}) The scattering matrix $S_q$ is replaced by a {\it pair}
of bonds (we refer to them as q-bonds), installed between the
forbidden regions $A_{n,m}$ and $A_{n\pm1,m\pm1}$, i.e., diagonal
bonds in Fig. \ref{pqmodel}. Both q-bonds are absent, Fig.
\ref{pqbonds}a, if the node deflects only to the left. Probability
of this realization is $P_a=(1-q)^2$, as follows from Eq.
(\ref{disq}). Deflection only to the right corresponds to two
crossed q-bonds present, Fig. \ref{pqbonds}b. This happens with
probability $P_b=q^2$. The situation when right-diagonal q-bond is
present while the left-diagonal q-bond is absent corresponds to
the scattering scenario in Fig. \ref{pqbonds}c. The opposite
scattering scenario, Fig. \ref{pqbonds}d, translates into
left-diagonal q-bond present and right-diagonal q-bond absent. The
two latter bond configurations have equal probabilities,
$P_c=P_d=q(1-q)$.
%%%%%%%%%%%%%%%%%%%
\begin{figure}[t]
\centerline{\includegraphics[width=60mm,angle=0,clip]{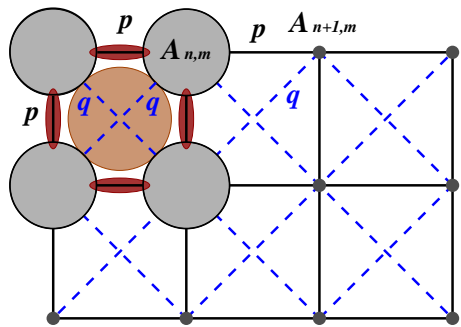}}
\caption{(Color online) Limit of strong disorder. The centers of
forbidden regions, $A_{n,m}$ and $A_{n,m-1}$, are connected by the
p- bond, while the centers of forbidden regions, $A_{n-1,m-1}$ and
$A_{n,m}$, are connected by a q- bond. The delocalization
transition corresponds to the percolation threshold on the lattice
consisting of p- and q- bonds.}
 \label{bondperc}
\end{figure}
%%%%%%%%%%%%%%%%%%

Quantum-mechanical delocalization transition in the limit of
strong disorder corresponds to percolation over p- and q-bonds,
see Fig. \ref{bondperc}. At the threshold of percolation p- and
q-bonds form an infinite cluster. At the same point, the waves
propagating along the links in both directions and scattered at
the links and at the nodes form an edge state. Threshold $(q,p)$
values lie on a critical line of transitions on a $q-p$ plain.
Crucial for us is the relation between the points of this line and
the positions of quantum delocalization transitions. In this
regard it is important to relate the quantum matrix $S_q$ to the
matrices describing the different classical scenarios shown in
Fig. \ref{pqbonds}. Setting $q=0$ we get
\begin{equation}
\label{Matrixa} S_a=\left(
\begin{array}{cccc}
0&-1&0&0\\
0&0&1&0\\
0&0&0&1\\
1&0&0&0
\end{array}\right),
\end{equation}
which describes the scattering in Fig. \ref{pqbonds}a. Scattering
scenario in Fig. \ref{pqbonds}b is described by the matrix
\begin{equation}
\label{Matrixb} S_b=\left(
\begin{array}{cccc}
0&0&0&-1\\
1&0&0&0\\
0&-1&0&0\\
0&0&-1&0
\end{array}\right),
\end{equation}
which emerges upon setting $q=1$ in Eq. (\ref{Matrixq}). To get
the matrix
\begin{equation}
\label{Matrixc} S_c=\left(
\begin{array}{cccc}
0&0&0&-1\\
0&0&1&0\\
0&-1&0&0\\
1&0&0&0
\end{array}\right),
\end{equation}
one has to set $q=0$ in the first and third columns, and $q=1$ in
the second and fourth columns. Similarly, the matrix
\begin{equation}
\label{Matrixd} S_d=\left(
\begin{array}{cccc}
0&-1&0&0\\
1&0&0&0\\
0&0&0&1\\
0&0&-1&0
\end{array}\right),
\end{equation}
corresponding to Fig. \ref{pqbonds}d emerges upon setting $q=1$ in
the first and third columns, and $q=0$ in the second and fourth
columns.

Matrices $S_a$ and $S_b$ provide non-zero Hall resistances,
$R_H^a=-1$, $R_H^b=1$, while for $S_c$ and $S_d$ we have
$R_H^c=R_H^d=0$, i.e., the Hall resistances are zero. The net Hall
resistance is thus determined by
\begin{equation}
\label{NHR}
P_aR_H^a+P_bR_H^b=P_b-P_a,
\end{equation}
which should be proportional to the magnetic field, $(1/2-q)$. The
other relations between the probabilities of different scattering
scenarios are normalization, $P_a+P_b+P_c+P_d=1$, and obvious
symmetry, $P_c=P_d$. These relations do not fix all probabilities
uniquely. There is a profound physical reason for this ambiguity.
Indeed, the net Hall resistance can be zero even if nodes locally
deflect either to the left or to the right provided that
$P_a=P_b$. This corresponds to the situation when a random
magnetic field with zero average acts on electron, so that the
time reversal symmetry is broken even in the absence of an
external field. Such situation is generic for composite fermions,
as was discussed in the Introduction.

In addition to the probability assignment
\begin{equation}
\label{newmod}
P_a=(1-q)^2,\quad P_b=q^2,\quad P_c=P_d=q(1-q),
\end{equation}
dictated by Eq. (\ref{disq}) and described above, one can choose,
e.g.,
\begin{equation}
\label{oldmod}
 P_a=1-q,\quad P_b=q, \quad P_c=P_d=0,
\end{equation}
when the electron scatters only to the left or only to the right
from all incident channels. Obviously, for the latter assignment
the magnitude of the random magnetic field is stronger than for
assignment Eq. (\ref{newmod}). Finally, the physical situation
when the time reversal symmetry is preserved in zero external
magnetic field corresponds to
\begin{equation}
\label{newestmod} P_a=1-2q,\quad P_b=0,\quad P_c=P_d=q.
\end{equation}
Three variants, Eqs. (\ref{newmod}), (\ref{oldmod}), and
(\ref{newestmod}), define three different percolation models,
which we denote as ${\cal A}$, ${\cal B}$, and ${\cal C}$,
respectively. Results of numerical simulations of these models are
reported in the next section.

In conclusion of the present section we would like to draw a
contrast between the classical limits of $4\times4$ scattering
matrix $S_q$ and of $2\times2$ scattering matrix $S_p$,
Eq.~(\ref{Matrixp}). Unlike the scattering matrix $S_q$, there is
no ambiguity in taking the strong-disorder limit of the $2\times2$
link matrix $S_p$ because this limit corresponds to the presence
or absence of a {\it single} bond. In this regard, note that, in
fully chiral network model by Chalker and Coddington \cite{CC},
scattering at the nodes is also described by a $2\times2$
scattering matrix. The limit of strong disorder corresponds to the
presence or absence of a single bond between the centers of the
squares $A_{n,m}$. Taking a strong-disorder limit in the
Chalker-Coddington model reduces the quantum problem to
conventional bond percolation on a square lattice. The position of
the percolation threshold and the quantum delocalization
transition certainly coincide, while the localization length in
the strong-disorder limit is smaller \cite{Kivelson93}.

\section{Simulation procedure and results}

In simulations performed, disorder realizations correspond to
presence or absence of p- and q-bonds. In each realization,
probabilities of p- and q-bonds are specified by the rules
formulated above. Convention for the p-bonds, connecting
counterpropagating links of $n+m$-odd and $n+m$-even sublattices
is the same for all three models. Conventions for q-bonds are
different for the models ${\cal A}$, ${\cal B}$, and ${\cal C}$.
These conventions are specified by Eqs. (\ref{newmod}),
(\ref{oldmod}), and (\ref{newestmod}), respectively. The main
peculiarity of the simulations that complicates the trajectories
stems from arrangement of pairs of q- bonds at the nodes. Namely,
for different directions of approach to the given node the
outcomes of passage are {\it correlated}. These correlations are
illustrated in Fig. \ref{bondperc}. The models ${\cal A}$, ${\cal
B}$, and ${\cal C}$ differ by the weights with which different
outcomes, $a$, $b$, $c$, or $d$, Fig. \ref{bondperc}, are allowed.

The size, $L$, of the samples used ranged between $500$ and
$10000$, where our unit of distance is half a link. For the
largest system, we average over $10^6$ disorder realizations. This
number increases with decreasing size so that we keep a roughly
constant CPU effort per size. To locate the position of the
percolation threshold we searched for trajectories connecting two
opposite faces of a square sample (periodic boundary conditions
were imposed in the perpendicular direction). As in Ref.
\onlinecite{ChalkOrtSom}, for a given realization, the
two-terminal conductance between the opposite open faces was
identified with the number of such spanning trajectories.
Different disorder realizations generate the conductivity
distribution with average, $\sigma(q,p,L)$.

\subsection{Phase diagrams}

To determine the critical boundary for each of the three models
considered, $p_{\cal A}(q)$, $p_{\cal B}(q)$, and $p_{\cal C}(q)$,
we select a set of values of the turning probability $q$ and then
scan for many values of the probability $p$. For each size $L$, we
represent the conductance as a function of $p$ on a logarithmic
scale and fit the points near the maximum of the conductance by a
Gaussian. We then plot the position of the peak as a function of
$L^{-1}$ and extrapolate to infinite size. We found empirically
that this fitting procedure is of high quality for the three
models.

\begin{figure}[th]
\centerline{\includegraphics[width=.48\textwidth]{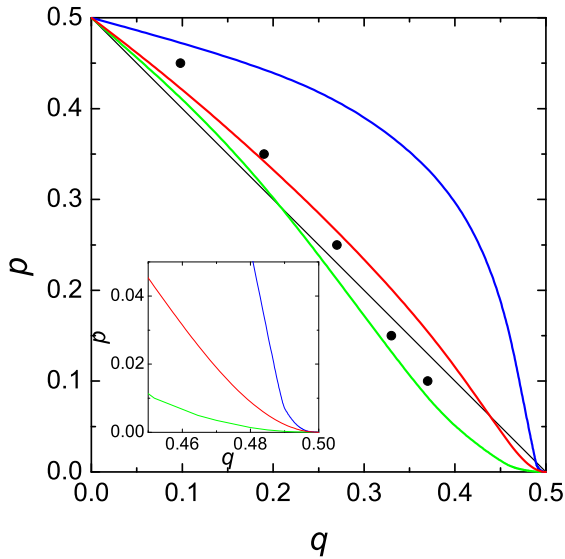}}
\caption{(Color online) Critical lines for the three model
considered: ${\cal A}$ (red, middle curve), ${\cal B}$ (blue,
upper curve) and ${\cal C}$ (green, lower curve). The dots are the
results of quantum simulations \cite{our}. Inset shows the same
three critical curves at at higher scale near the point
$(1/2,0)$.}
 \label{phasediag}
\end{figure}

In Fig. \ref{phasediag} we represent the critical lines obtained
for the three models: the upper curve corresponds to model ${\cal
B}$, the curve in the middle to model ${\cal A}$, and the lower
curve to model ${\cal C}$. The straight line corresponds to
$p=1/2-q$. The solid dots are the results of the quantum
simulations in Ref. \onlinecite{our}. The inset shows the same
critical lines as in the main panel at larger scale near the point
$(1/2,0)$.
%%%%%%%%%%%%%%%%%%%
%%%%%%%%%%%%%%%%%%

\begin{figure}[th]
\centerline{\includegraphics[width=.48\textwidth]{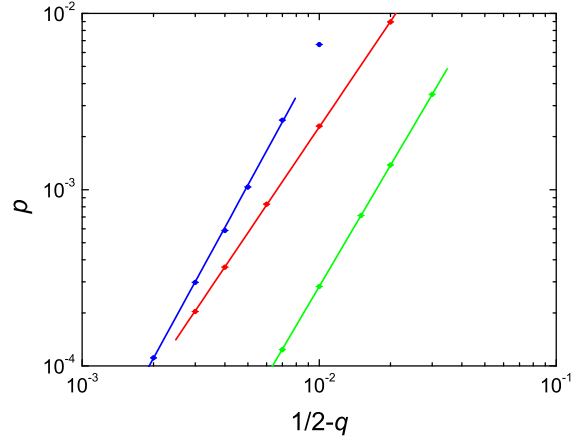}}
\caption{(Color online) Critical lines for the three model
considered ${\cal A}$ (red, middle curve), ${\cal B}$ (blue, upper
curve) and ${\cal C}$ (green, lower curve) on a double logarithmic
scale near the point $(1/2,0)$.}
 \label{logscale}
\end{figure}

We note that the curves tend to the point $(1/2,0)$ as a power law
with power {\it higher than linear}. To gain further insight, we
analyze in detail the shape of the phase boundary at small
probabilities $p$. In this regime $q_{\rm c}$ is close to $1/2$
and we expect a relation of the form
\begin{equation}\label{gamma}
p\propto \left|q-\frac{1}{2}\right|^\gamma .
\end{equation}
In Fig. \ref{logscale} we show $p$ versus $q-1/2$,  on a double
logarithmic scale, for the three models ${\cal A}$ (middle set of
points), ${\cal B}$ (upper set) and ${\cal C}$ (lower set). The
straight lines are  linear fits to the corresponding points. Their
slopes are $\gamma_{\cal A}=1.994 \pm 0.001$, $\gamma_{\cal
B}=2.464 \pm 0.001$ and $\gamma_{\cal C}=2.286 \pm 0.001$. The
errors quoted are the statistical errors; systematic errors are
also present, since it is impossible to include all finite-size
effects.
%%%%%%%%%%%%%%%%%%%
%%%%%%%%%%%%%%%%%%

\subsection{Conventional percolation behavior  away from $q=1/2$.}

Quantum simulations in Ref. \onlinecite{our} demonstrated that the
delocalization transition along the boundary, $p(q)$, belongs to
the quantum Hall universality class. In particular, it was
demonstrated that for the first three black dots in Fig.
\ref{phasediag} which correspond to $q<0.3$, the critical exponent
is close to $7/3$. Introducing strong disorder suppresses the
quantum interference. We expect that quantum percolation at a
given $(q,p)$ reduces to classical percolation and the critical
exponent $\nu=7/3$ is replaced by its classical value $\nu=4/3$.
For the Chalker-Coddington model, in which the position of
delocalization is fixed at average value of the disorder
potential, the crossover from $7/3$ to $4/3$ was tested in Ref.
\onlinecite{Kivelson93}. In this section we demonstrate that, {\it
away from the point} $p=0$, $q=1/2$, the $p(q)$ boundary
established above indeed corresponds to the divergence of the
localization length with exponent $\nu=4/3$.

The determination of the critical exponent is based on the fact
that near $(q,p)=(q_c,p_c)$ the conductance is a function of a
single argument, $(p-p_c)L^{1/\nu}$ (vertical scan) or
$(q-q_c)L^{1/\nu}$ (horizontal scan). Exactly at $(q_c,p_c)$ the
conductivity assumes the universal value,
$\sigma_0=0.361404\ldots$ found by Cardy \cite{cardy02}, which
should be the same for the entire boundary.
%The critical behavior of the three models away from the line
%$q=1/2$ is that of percolation. The correlation length exponent is
%$\nu=4/3$ and the conductivity tends to the exact value found by
%Cardy \cite{cardy02} (see also \cite{john}) $0.361404\ldots$.

\begin{figure}[th]
\centerline{\includegraphics[width=.48\textwidth]{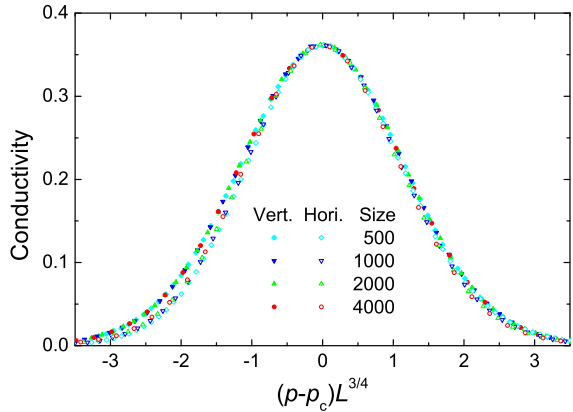}}
\caption{(Color online) Scaled conductivity as a function of the
probability difference with the critical point multiplied by
$L^{3/4}$. The solid symbols correspond to vertical scans and the
empty symbols to horizontal scans crossing the critical point
$(0.3,0.23392)$ of model ${\cal A}$. The lateral sample sizes are:
500 (cyan diamonds), 1000 (blue down triangles), 2000 (green up
triangles) and 4000 (red circles).}
 \label{percolation}
\end{figure}

The scaling analysis was performed for all three models. In Fig.
\ref{percolation} we present results for the model ${\cal A}$ at a
particular critical point $(0.3,0.23392)$. Overall, the scaling
confirms that $\nu=4/3$ both for vertical and horizontal scans.
Particular feature about the scaling data is that the widths of
scaling functions are slightly different for the vertical and
horizontal scans. We have also found that there is a small size
effect precisely at the boundary which is well described by the
expression
\begin{equation}\label{finite}
\sigma_L=\sigma_0+ \frac{a}{L^{3/4}},
\end{equation}
where $a$ is a constant.
%We have checked that this is indeed so for the three models and
%here we present results for the model ${\cal A}$ in the critical
%point $(0.3,0.23392)$. The maximum value of the conductance as a
%function of $p$ for fixed $q$ (or alternatively as a function of
%$q$ for fixed $p$) presents small size effects, which are well
%represented by the expression
%where $a$ is a constant and $\sigma_0$ is the macroscopic value of
%the conductivity. For the point $(0.3,0.23392)$, for example, we
%found $\sigma_0=0.3604\pm 0.0010$ in agreement with the exact
%value for percolation. Everywhere along the critical lines the
%conductivity tends, within errors, to the value for percolation,
%except right at the point $(1/2,0)$ where it tends to twice this
%value.
%In Fig. \ref{percolation} we plot the conductivity scaled by the
%factor $\sigma_0/\sigma_L$ for several sizes. There are two set of
%data, in one case we take $q=0.3$ and plot $(p-0.23392)L^{1/\nu}$
%(solid symbols) and in the other we take $p=0.23392$ and plot
%$(q-0.3)L^{1/\nu}$ (empty symbols). In both cases we take for
%$\nu$ the percolation exponent for the correlation length
%$\nu=4/3$. We note that each set of data collapse  to a different
%curve. The proportionality factor depends on the scanning
%direction, but the scaling exponent is  the same, coinciding with
%the value for percolation.
%%%%%%%%%%%%%%%%%%%
%%%%%%%%%%%%%%%%%%

\subsection{Behavior of the localization length at zero field: models $\cal A$ and $\cal C$.}

We now turn to the behavior of the localization length, $\xi$, at
zero magnetic field, $q=1/2$. For each model, the dependence
$\xi(p)$ at small $p$ is determined by a peculiar behavior of the
corresponding delocalization boundary established in subsection A.
It is also very important that $\xi(p)$ is equally affected by the
second, complementary, boundary in the domain $q>1/2$, which is
the mirror image of the boundary in Fig. \ref{phasediag}.

For moderate $p$, when the boundaries were straight lines
\cite{our}, the presence of the second boundary leads to the
enhancement of $\xi$. On the contrary, we will see that in the
limit $p\rightarrow0$, the fact that both boundaries for a given
model are almost horizontal leads to a shortening of $\xi$.

%%%%%%%%%%%%%%%%%%%
\begin{figure}[th]
\centerline{\includegraphics[width=.48\textwidth]{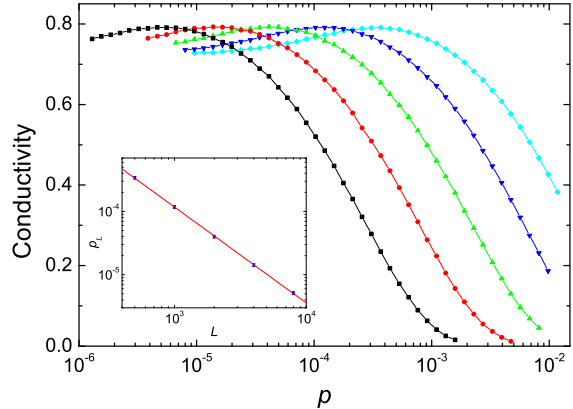}}
\caption{(Color online) Conductivity as a function of $p$ for
model ${\cal A}$ along the line $q=1/2$. The lateral sample sizes
are: 500 (cyan diamonds), 1000 (blue down triangles), 2000 (green
up triangles), 4000 (red circles) and 8000 (black squares). Inset
shows the position of the peaks as a function of $L$ on a double
logarithmic scale.}
 \label{rawA}
\end{figure}
%%%%%%%%%%%%%%%%%%

\begin{figure}[th]
\centerline{\includegraphics[width=.48\textwidth]{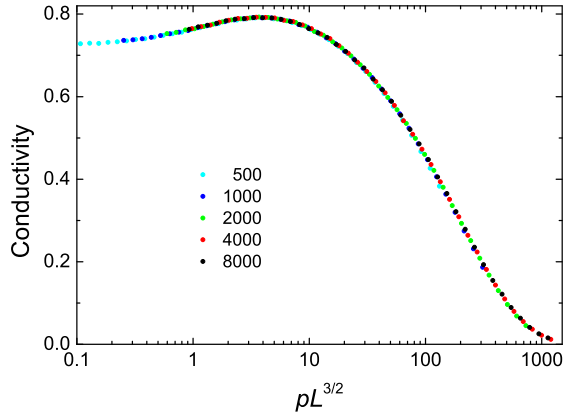}}
\caption{(Color online) Conductivity as a function of $pL^{3/2}$
on a logarithmic scale for  model ${\cal A}$ along the line
$q=1/2$. The sample sizes are: 500 (cyan), 1000 (blue), 2000
(green), 4000 (red) and 8000 (black).}
 \label{OverlapA}
\end{figure}

In Fig. \ref{rawA} we plot the conductivity for model $\cal A$ as
a function of $p$ for $q=1/2$ and for several values of the system
size: 500 (diamonds), 1000 (down triangles), 2000 (up triangles),
4000 (circles) and 8000 (squares). It is seen that the curves
$\sigma(p)$ for different system sizes have similar shapes and are
even-spaced along the logarithmic horizontal axis. Thus we expect
scaling and behavior $\xi\sim p^{-\nu_{\cal A}}$ as a consequence.
A practical procedure to infer $\nu_{\cal A}$ from the data in
Fig. \ref{rawA} is based on the dependence, $p_L$ versus $L$,
where $p_L$ is the position of the maximum of the conductivity for
a given $L$. In the inset of Fig. \ref{rawA} we plot $p_L(L)$ in a
double logarithmic scale. We see that $p_L$ follows the dependence
\begin{equation}
p_L=bL^{-\beta}, \label{pL}
\end{equation}
where $b$ and $\beta$ are model-dependent constants. For model
$\cal A$, we found $\beta_{\cal A}=1.51\pm 0.02$ and $b=4.1\pm
0.8$. Eq. (\ref{pL}) and the fact that $\beta_{\cal A}$ is very
close to $3/2$ suggest that, to achieve scaling, the data in Fig.
\ref{rawA} should be replotted versus $pL^{3/2}$. The result of
this replotting is shown in Fig. \ref{OverlapA}. It is seen that
the overlap is excellent, yielding the critical exponent
$\nu_{\cal A}=2/3$. This should be contrasted to the behavior
$\xi(p)\propto(p-p_c)^{-4/3}$ at {\it any} non-zero $p_c$. We
conclude that at $p_c=0$ the divergence of the localization length
with $p$ is much slower.
%%%%%%%%%%%%%%%%%%%
%%%%%%%%%%%%%%%%%%

\begin{figure}[th]
\centerline{\includegraphics[width=.48\textwidth]{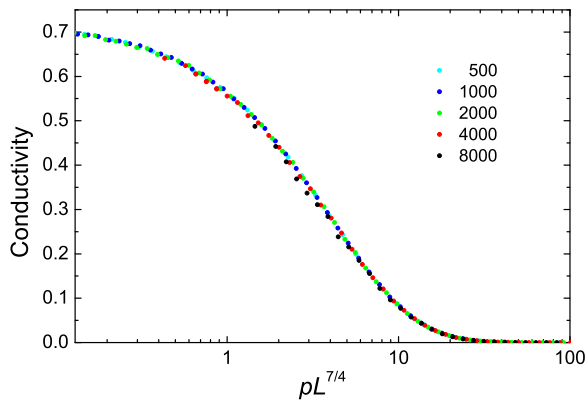}}
\caption{(Color online) Conductivity as a function of $pL^{7/4}$
on a logarithmic scale for  model ${\cal C}$ along the line
$q=1/2$. The sample sizes are: 500 (cyan), 1000 (blue), 2000
(green), 4000 (red) and 8000 (black).}
 \label{OverlapC}
\end{figure}

Finally, for the model $\cal C$ the plots $\sigma(p)$ as a
function of $p$ do not exhibit maxima. As shown in Fig.
\ref{OverlapC}, where we plot $\sigma(p)$ versus $pL^{7/4}$, a
very good overlap is achieved for $\nu_{\cal C} =4/7$.
%%%%%%%%%%%%%%%%%%%
%%%%%%%%%%%%%%%%%%

In conclusion of this subsection we note that the conductivity for
all three models tends to $2\sigma_0=0.722808...$ as
$p\rightarrow0$.
%maximal conductivity in Figs. \ref{OverlapA}-\ref{OverlapC} is
%$0.7-0.8$, which is approximately twice the value of $\sigma_0$.
The reason is that the value $\sigma=\sigma_0$ at the threshold is
the property of a {\it single} critical point \cite{cardy02}. By
contrast, in our case {\it two} critical lines merge at the point,
$q=1/2$, $p=0$.

\subsection{Behavior of the localization length at zero field: model $\cal B$.}

\begin{figure}[th]
\centerline{\includegraphics[width=.48\textwidth]{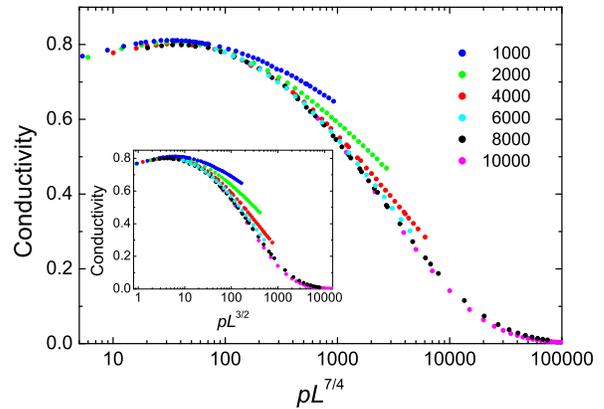}}
\caption{(Color online) Conductivity as a function of $pL^{7/4}$
along the line q = 1/2 on a logarithmic scale for model $\cal B$.
The sample sizes are: 1000 (blue), 2000 (green), 4000 (red), 6000
(cyan), 8000 (black) and 10000 (magenta). Inset: the same as main
plot with low-p data included; conductivity is plotted versus
$pL^{3/2}$.}
%Conductivity as a function of $(p-p_L)L^{15/8}$ along the line
%$q=1/2$ on a logarithmic scale for model ${\cal B}$. The sample
%sizes are: 1000 (blue), 2000 (green), 4000 (red), 6000 (cyan),
%8000 (black) and 10000 (magenta).}
 \label{OverlapB}
\end{figure}

Scaling analysis of the data for model $\cal B$ reveals slightly
different behaviors for the domains of "moderate" $p> 10^{-4}$ and
"truly critical" $p<10^{-4}$. For the first domain, from the
position of peaks we find $\beta_{\cal B}=1.77\pm 0.04$. This
suggests that $\nu_{\cal B}=1/\beta_{\cal B}\approx 4/7$. Note
however that replotting the conductivity versus $pL^{7/4}$, see
Fig. \ref{OverlapB}, does not lead to overlap as good as for the
model $\cal A$. Moreover, in the second domain $p<10^{-4}$ a good
overlap is achieved for the exponent $2/3$, i.e., the same as in
the model $\cal A$. This is illustrated in the inset of Fig.
\ref{OverlapB}. This indicates that for the model $\cal B$ the
true critical region is quite narrow. Such a delicate behavior of
$\xi(p)$ for the model $\cal B$ might indicate that the critical
boundary $p(q)$ in this model also changes the behavior in the
truly critical region $p\lesssim 10^{-4}$.

%%%%%%%%%%%%%%%%%%%
%%%%%%%%%%%%%%%%%%

\section{Discussion}

\subsection{Position of boundaries}

It is seen from Fig. \ref{phasediag} that the boundaries, $p_{\cal
A}(q)$ and $p_{\cal C}(q)$, almost coincide in the entire domain,
$0<q<1/2$. Overall, these boundaries are in agreement with the
results of quantum simulation Ref. \onlinecite{our} shown with
black dots. It is also seen that the boundary, $p_{\cal B}(q)$,
goes significantly higher. In particular, at $q=0.25$, $p_{\cal
B}$ exceeds $p_{\cal C}$ almost twice. On the physical level, this
means that, for a given average magnetic field, the formation of
edge states requires a longer zero-field mean free path, $l$, for
model ${\cal C}$ than for model ${\cal B}$. In other words,
formation of edge states happens easier when a random magnetic
field is present. To gain a physical insight why this is so,
consider electron motion in random magnetic field. Local value of
the field changes its sign in space, while the average field,
$(1/2-q)$, is much smaller than the absolute value of the local
field. Then electron trajectories are either circles inside the
regions where the field maintains its sign, or snake states,
propagating along the boundaries of these regions, i.e., along the
contours with zero local field. Then it is apparent that a weak
disorder does not affect this picture. If, on the other hand, the
magnetic field, $(1/2-q)$, is uniform, electron trajectories are
big circles. Then a weak disorder will have a strong effect by
deflecting electron before it completes a circle. The above two
situations correspond to the models $\cal B$ and $\cal C$,
respectively, and explain why $p_{\cal B}(q)> p_{\cal C}(q)$. In
model $\cal A$, random component of magnetic field is present, but
is weaker than in model $\cal B$. In this regard, the fact that
the boundary  $p_{\cal A}(q)$ lies between  $p_{\cal B}(q)$ and
$p_{\cal C}(q)$, also finds its explanation.

Although quantum simulations are impossible in the domain $p\sim
0.01$ five "quantum" data points in Fig. \ref{phasediag} follow
$p_{\cal C}(q)$ within the accuracy of quantum simulations.
%Our indirect method to establish the quantum boundary for smaller
%$p$ is based on the conjecture that this boundary follows $p_{\cal
%C}(q)$ down to $p=0$.
Then the confirmation of the scaling theory Eqs. (\ref{pru1}),
(\ref{pru2}) would be the linearity of the percolation boundary at
$q\rightarrow 1/2$, see Eq. (\ref{predbound}).

The most important outcome of the present simulation is the inset
in Fig. \ref{phasediag}. It is seen that at really small $p\sim
0.01$ and $q$ close to $1/2$ the behaviors of all three boundaries
changes dramatically compared to their "bodies", namely, they
become almost horizontal. All three boundaries have the form
$p\sim(1/2-q)^\gamma$ with $\gamma\gtrsim2$. This is in stark
contrast to the prediction of scaling theory Eq.
(\ref{predbound}), which corresponds to $\gamma=1$. In other
words, percolation results suggest that instead of the condition
$\sigma_{xy}=1/2$, the delocalization boundary is described by
\begin{equation}
\label{boundary} \sigma_{xy}\sim\sigma_{xx}^{1-\frac1\gamma}.
\end{equation}
The latter condition can be also cast in the form Eq.
(\ref{newscale}) with $\kappa=1-1/\gamma$. We note that the
crossover from $\sigma_{xy}=1/2$ to Eq. (\ref{boundary}) takes
place at large $\sigma_{xx}\sim 10$. In terms of the flow diagram
of the quantum Hall effect \cite{Khmelnitskii} this means that the
upper part of the vertical flow line is bent to the right, as it
is illustrated in Fig. \ref{levit}b.

\subsection{Semi-analytical consideration}

To specify the distinct behavior of percolation boundaries in
vanishing average magnetic field they are plotted in
Fig.~\ref{logscale} in the log-log scale. From the slopes we
deduce the values, $\gamma_{\cal A}=1.994$, $\gamma_{\cal
B}=2.464$, and $\gamma_{\cal C}=2.286$. To get a feeling why all
$\gamma$-values are close to $2$, below we present some
semi-analytical arguments.
%Below we present a semi-analytical derivation
%of $\gamma_{\cal A}$ and $\gamma_{\cal B}$.
We first turn to Fig. \ref{bondperc} and set $p=0$. Then the
lattice breaks into two quadratic sublattices with $n+m$ even and
$n+m$ odd, which are completely disconnected. None of them
percolates if $q$, the percentage of bonds present in each
sublattice, is less than $1/2$. Finite $p=p_c(q)$ allows
percolation for $q<1/2$ since p-bonds couple clusters from
different sublattices. It is apparent that coupling of clusters by
p-bonds is relevant if the typical distance, $1/\sqrt{p}$, between
two p-bonds become smaller than the localization length,
$\xi(q)=(1/2-q)^{-4/3}$. This yield a constrain that $\gamma<8/3$.
This constrain is insensitive to the mutual correlations of
q-bonds on the two sublattices. In fact, this correlation is
absent in model $\cal A$. Indeed, as follows from Eq.
(\ref{newmod}), at $q=1/2$ for model $\cal A$ we have
$P_a=P_b=P_c=P_d=1/4$. By contrast, for model $\cal B$ the
probabilities at $q=1/2$ are $P_a=P_b=1/2$, $P_c=P_d=0$. This
suggests that q-bonds on two sublattices are strongly (and
positively) correlated. Namely, if there is a q-bond connecting
two $n+m$ even plaquettes at a given node, then there must be a
q-bond connecting $n+m$ odd plaquettes at the same node. On the
other hand, the correlation of q-bonds at a node in model $\cal C$
is negative: presence of one q-bond excludes the presence of the
other. This is apparent from Fig. \ref{pqbonds}.
%%%%%%%%%%%%%%%%%%
\begin{figure}[t]
\centerline{\includegraphics[width=50mm,angle=0,clip]{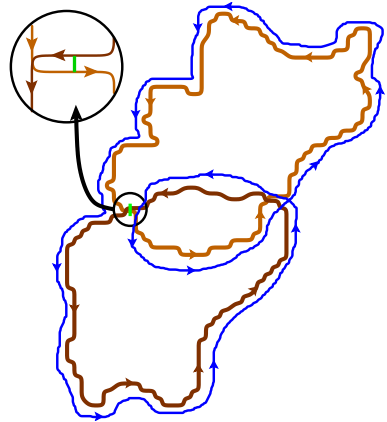}}
\caption{(Color online) Vicinity of the point $p=0$, $q=1/2$, of
the phase diagram Fig. \ref{logscale}.  A q-cluster of $n+m$ odd
sublattice (upper) and a q-cluster of $n+m$ even sublattice
(lower) overlap. A joint trajectory (thin blue line) is formed
upon installing of a {\it single} p-bond. The blowup illustrates
hybridization of trajectories on a microscopic level.}
\label{clusters}
\end{figure}
%%%%%%%%%%%%%%%%%%

As p-bonds are switched on, clusters include sites from both
sublattices, see Fig. \ref{clusters}. In Ref. \onlinecite{our}
where the model $\cal A$ was considered, it was argued that
$\gamma=1$. The argument was based on the following picture of the
cluster growth upon increasing $q$: it was assumed that critical
q-clusters on a given sublattice grow by getting connected via
additional q-bonds. Since the growth of clusters due to p-bonds
takes place by connecting critical clusters from different
sublattices, it was concluded that p- and q-bonds play equal roles
in the growth of clusters, which immediately leads to $\gamma=1$.
Present simulations suggest that in the close proximity of $q=1/2$
this picture fails, and the role of p- and q-bonds in approaching
the percolation threshold is completely different, namely, the
growth due to p-bonds is more efficient. This growth proceeds by
p-bonds connecting the {\it hulls} of critical clusters from two
sublattices, as illustrated in Fig. \ref{clusters}. For a given
$q$, the length of the hull is
\begin{equation}
\label{hull} {\cal
L}(q)=\xi^{7/4}(q)=\left(\frac12-q\right)^{-7/3}.
\end{equation}
For the model $\cal A$, to achieve a percolation by adding p-bonds
one should take into account that the hulls on two sublattices are
uncorrelated. Then, a p-bond with one end on a hull from even
sublattice will have the other end on the hull from odd sublattice
with probability ${\cal L}/\xi^2$. As a result, the percolation
condition reads
\begin{equation}
\label{moreperhull} ({\cal L}/\xi^2)(p{\cal L})=1,
\end{equation}
where the second factor is the probability that there is at least
one p-bond with one end on the critical hull from, say, odd
sublattice. Eq. (\ref{moreperhull}) yields $\gamma_{\cal A}=2$,
which coincides with the simulation result.

Correlation of q-bonds at the nodes for models $\cal B$ and $\cal
C$ leads to conclusion that corresponding critical clusters in two
sublattices are also correlated. One consequence of this
correlation is that it takes less p-bonds than in model $\cal A$
to connect critical hulls from two sublattices. As a result,
$\gamma_{\cal B}$, $\gamma_{\cal C}\geq\gamma_{\cal A}=2$. On the
other hand, the picture of percolation by connecting the critical
clusters imposes the upper boundary $\gamma\leq7/3$ for both
models $\cal B$ and $\cal C$. This follows from the condition that
there should be at least one p-bond per critical hull, i.e.,
$p{\cal L}\geq1$. Note that the constraint $\gamma\leq7/3$ is
stricter than the constraint $\gamma<8/3$, established above.

Beyond the estimate, $2\leq\gamma_{\cal B}$, $\gamma_{\cal
C}\leq7/3$ we cannot come up with more accurate analytical values
for these indices. We are not even able to establish which of them
is bigger. This is because strong correlation between the hulls in
both models $\cal B$ and $\cal C$ simplifies connectivity upon
switching on p-bonds. On the other hand, this correlation prevents
the expansion of the resulting cluster.

As it was established in the previous section, in the domain of
$p\lesssim 10^{-4}$, behavior of $\xi(p)$ at $q=1/2$ in the model
$\cal B$ exhibits crossover from the critical exponent, $4/7$, to
$2/3$. To relate this peculiar behavior with the shape of the
percolation boundary $p_{\cal B}(q)=(1/2-q)^{\gamma_{\cal B}}$, we
invoke the argument of Ref. \onlinecite{KagaChalker} which, in
application to the p-q model, goes as follows. If the divergence
of $\xi$ at the point, $q=1/2$, $p=0$, is characterized by
$\xi\sim(1/2-q)^{-\nu_q}$ along the q- direction and $\xi\sim
p^{-\nu_p}$ along the p- direction, then the shape of the critical
boundary is $p\sim (1/2-q)^{\nu_q/\nu_p}$, i.e.,
$\gamma=\nu_q/\nu_p$. Following this argument, crossover in the
model $\cal B$ from $\nu_p=\nu_{\cal B}=4/7$ to $\nu_{\cal B}=2/3$
suggests that $\gamma_{\cal B}$ and $\gamma_{\cal A}$ merge in
truly critical region.
%causing the violation of the relation
%$\gamma_{\cal B}>\gamma_{\cal C}>\gamma_{\cal A}$.

Overall, our numerical results suggest that in the truly critical
domain, where $\gamma_{\cal A}\approx\gamma_{\cal B}\approx2$ and
$\gamma_{\cal C}\approx 7/3$, the divergence of $\xi(p)$ for all
three models is well described by the relation
\begin{equation}
\label{allthree} \xi\sim p^{-4/(3\gamma)}.
\end{equation}
With regard to the argument of Ref. \onlinecite{KagaChalker} this
means that, for all three models, the exponent $\nu_q$ is equal to
$4/3$, i.e., the same as for $q$ away from $1/2$.

\subsection{Relation to Ref. \onlinecite{ChalkOrtSom}}

In Ref. \onlinecite{ChalkOrtSom}, spin quantum Hall effect in
bilayer and trilayer systems was studied numerically, in order to
trace the emergence of macroscopic metallic phase upon adding the
third dimension \cite{OrtSomChal}. The authors made use of the
fact that in a strictly 2D system there is a mapping between the
spin quantum Hall transition and classical bond percolation
\cite{GruzRead, BeaCarChal}. For bilayer systems the corresponding
classical percolation is bond percolation on each layer (bonds
connect the centers of plaquettes), complemented with the
possibility to switch layers with a probability, $p_1$, while
passing each side of each plaquette. Physically, in spin quantum
Hall effect, an electron travels on each layer of the network in
the same direction. In our consideration of the weak-field quantum
Hall effect, an electron stays within a plane but each link of the
square lattice represents two counterpropagating channels. For
this reason there is a mapping between the simulation in Ref.
\onlinecite{ChalkOrtSom} and treatment of the model $\cal A$ in
the present paper. Namely, $p_1$ in Ref. \onlinecite{ChalkOrtSom}
should be identified with the backscattering probability $p$ in
the present paper, while the probability that the given bond is
present in Ref. \onlinecite{ChalkOrtSom}, $p$, should be
identified with our parameter $q$. Due to this mapping, critical
behavior, $p_1(p)$ for small $p_1$ in Ref.
\onlinecite{ChalkOrtSom}, is the same as the behavior of critical
line, $p\propto(1/2-q)^{\gamma_{\cal A}}$, for small $p$ in our
model $\cal A$. Also, the above semi-analytical calculation of
$\gamma_{\cal A}$ is the same as proposed in Ref.
\onlinecite{ChalkOrtSom}. However, it should be noted that mapping
between the network of Ref. \onlinecite{ChalkOrtSom} and model
$\cal A$ applies only for small $p_1$. For larger $p_1$ the
position of the boundary in Ref. \onlinecite{ChalkOrtSom} differs
dramatically from that of the model $\cal A$.

\subsection{Phases with higher $\sigma_{xy}$}

In fact, scaling theory predicts that {\it all} Landau levels,
$n$, in Eq. (\ref{positions}), eventually levitate to
$E_n\rightarrow\infty$, as the magnetic field is lowered. Our
simulations do not capture low-field transitions for $n\geq1$.
This is because we restricted our consideration to network with
one channel per link. Within this description we were able to
capture the Drude conductivity tensor of electron gas in a weak
field, and weak localization effects. On the other hand, this
description does not allow, in principle, to capture the phases
with quantized $\sigma_{xy}$ higher than $1$.

As a final remark, we can underscore the difference of the scaling
theory and our results as follows. The scaling theory predicts
that electron gas experiences a delocalization transition in a
magnetic field at which flux into the area $l^2$ is of the order
of the flux quantum, $\Phi_0$. We find that, in the limit of
$k_{\scriptscriptstyle F}l>10$, the interplay of orbital and phase
actions of magnetic field causes the transition when this flux is
much larger than $\Phi_0$.

\section{Acknowledgements}
We acknowledge the hospitality of KITP Santa Barbara where this
project was initiated. V. V. M. and M. E. R. acknowledge the
support of the Grants DOE No. DE-FG02-06ER46313 and BSF No.
2006201, and useful discussions with I. Gruzberg and V.
Kagalovsky. M. O. and A. M.  S. thank financial support from
Spanish DGI Grant No. FIS2009-13483, and from Fundacion Seneca,
Grant No. 08832/PI/08.

\end{document}